\documentclass[a4paper,12pt]{article}
 \usepackage[latin1]{inputenc}
\usepackage[T1]{fontenc}
\usepackage{amsmath}
\usepackage{stackrel}
\usepackage{amsfonts}
\usepackage{slashed}
\usepackage{amssymb}
\usepackage{color}
\usepackage{graphicx}
\usepackage{color}
\usepackage{dsfont}
\usepackage{upgreek}
\usepackage{soul}
\usepackage[normalem]{ulem}

\def\vF{\vec{F}} \def\vPhi{\vec{\Phi}} 
\def\vA{\vec{A}}

 \newcommand{\be}{\begin{equation}}
  \newcommand{\bb}{\begin{equation}}
	 \newcommand{\ee}{\end{equation}}
	 \newcommand{\ba}{\begin{eqnarray}}

		 \newcommand{\ea}{\end{eqnarray}}
		 
		   \newcommand{\bea}{\begin{eqnarray}}
			 \newcommand{\eea}{\end{eqnarray}}
\newcommand{\eqb}{\begin{eqnarray}}
\newcommand{\eqf}{\end{eqnarray}}

\begin{document} \title{Fermion zero modes in a $Z_2$ vortex background}
\author{ Gustavo Lozano$^a$, Azadeh Mohammadi$^b$, Fidel A. Schaposnik$^c$
~\\
~\\
{$^a\,$\it \small Departamento de F\'\i sica, FCEYN Universidad de Buenos Aires $\&$ IFIBA CONICET,}\\
{\it \small Pabell\'on 1 Ciudad Universitaria, 1428 Buenos Aires, Argentina}
\\
~\vspace{-3 mm}
\\
{$^b\,$\it \small Departamento de F\'\i sica, Universidade Federal da Campina Grande}\\
{\it \small   Caixa Postal 10071, 58429-900 Campina Grande, PB, Brazil}
\\
~\vspace{-3 mm}
\\
{$^c\,$\it \small Departamento de F\'\i sica, Universidad Nacional de La Plata/IFLP/CICBA}
\\{\it \small CC 67, 1900 La Plata, Argentina}
}

\date{\today}

\maketitle
\begin{abstract}
In this paper we study the zero energy solutions of the Dirac equation in the background of a $Z_2$ vortex of a non-Abelian gauge model with three charged scalar fields. We determine the number of the fermionic zero modes giving their explicit form for two specific Ansatze.
\end{abstract}
\section{Introduction}

The spectrum of Dirac like operators in the presence of topologically non-trivial backgrounds have attracted the attention of physicist  since the early work of Jackiw and Rebbi \cite{JRe} discussing the cases $d=1$  and $d=3$ soliton backgrounds (kinks and monopoles) as well as $d=4$  instanton backgrounds. In particular, 't Hooft found a solution of a notorious problem in high energy physics, the so called  U(1) problem in QCD \cite{thooft}-\cite{Co}  taking into account the contribution of   the Dirac operator zero modes in a topologically non-trivial gauge background of instanton configurations.

 Later on Jackiw and Rossi~\cite{JRO} considered the case in which the topological background is provided by vortex-like configurations and explicitly constructed the zero modes of the Dirac operator  in $d=2$ spatial dimensions. The  result suggested that also in  two-dimensional non-compact spaces the index theorem is valid, as was afterward proven in~\cite{We}. Interestingly enough, the Jackiw and Rossi zero modes can be chosen to be   eigenmodes of a particle conjugation operator and hence considered as Majorana zero modes (see \cite{Wil} and references therein).

The physical implications of zero modes are very surprising. Apart from their QCD application mentioned above, they are at the basis of charge fractionalization and cosmic string superconductivity, just to name some few examples (\cite{JRe},
\cite{Witten}-\cite{Polchinski}).

Concerning planar physics,  a more recent wave of interest started after the realization that Majorana quasiparticles can appear in some solid states systems, the topological superconductors, and they could play an important role in building topological protected qubits~\cite{review}.

As mentioned before, in $d=2$ dimensional systems, the existence of zero modes is linked to the presence of a vortex-like background. The original work of Jackiw and Rossi~\cite{JR} was concerned with zero modes of electrons moving in the background of a Nielsen-Olesen vortex. Many generalizations are possible. For instance the case in which the vortex background is the one arising in a non-Abelian theory was considered in~\cite{CL}-\cite{Shifman}. Zero modes for the case of  a Chern-Simons vortex background  were studied in \cite{GN}-\cite{LLM} and more recently  in the context of models having hidden sectors  \cite{LMS} that could be relevant in connection to superconductivity \cite{Sha}.

Recently a new type of $Z_2$ vortices  in non-Abelian gauge theories was presented in~\cite{CLS}. This type of configuration is a local generalization of magnetic vortices that appear in some triangular lattices of antiferromagnetic materials  \cite{KM}. It corresponds to a non-Abelian $SU(2)$ gauge theory with {\em three} scalar Higgs fields in the adjoint representation. We analyse in this paper, the existence of fermionic zero modes under such backgrounds by constructing them explicitly.

The paper is organized as follows: in section 2, we briefly review the $Z_2$ vortices  in non-Abelian gauge theories coupled to three scalar triplets \cite{CLS} that will be taken as a background of the Dirac equation defining the zero mode problem. Then in section 3 we introduce the Lagrangian for fermions minimally coupled to the non-Abelian gauge field background and also including a scalar-fermion coupling inspired in the one introduced in \cite{CL} for studying the zero mode problem in the background of  the $Z_N$ vortices discussed in \cite{deVS}. After proposing an axially symmetric Ansatz, we are able to decouple the gauge field thanks to the existence of a charge conjugation operator that reduces the zero mode equations to ordinary radial  differential equations in the scalar field background. Solving these equations we find the explicit form and number of the zero modes. We present in section 4 a summary of our results and a discussion of possible applications.

\section{The vortex background}
As a background for the Dirac fermion equation, we consider the vortex solutions found in  \cite{CLS} for a $SU(2)$ gauge theory coupled to three scalar fields in the adjoint representation.
The $2+1$ dimensional Lagrangian leading to vortex configurations reads
\be
L=-\frac{1}{4} \vF_{\mu \nu} \vF^{\mu \nu} +\frac{1}{2}D_\mu \vPhi_a D^\mu \vPhi_a-V(\vPhi_a)
\label{L}
\ee
Here  the gauge fields $A_\mu$  take values in the Lie algebra of $SU(2)$, $A_\mu = \vec A_\mu \cdot \vec\sigma /2$  while the scalars in the adjoint representation are written as $\Phi_a = \vec\Phi_a\cdot \vec \sigma/2$ ($a = 1,2,3$ and  $\vec\sigma$ are the Pauli matrices). Field strengths and covariant derivatives are defined as
\be
\vF_{\mu \nu}=\partial_\mu \vA_\nu -\partial_\nu \vA_\mu + e \vA_\mu \times \vA_\nu
\ee
\be
D_\mu \vPhi_a= \partial_\mu \vPhi_a + e \vA_\mu \times \vPhi_a
\label{ddd}
\ee
As for the potential, one has
\be
V(\vPhi_a)=\lambda_1 (\vPhi_1 \cdot \vPhi_1-\eta^2)^2 + \lambda_2 (\vPhi_2 \cdot \vPhi_2-\eta^2)^2+\lambda_3 (\vPhi_3 \cdot \vPhi_3-\eta^2)^2 + V_{mix}(\vPhi_a)
\ee
where
\be
V_{mix}(\vPhi_a)=\mu^2 (\vPhi_1 +\vPhi_2+\vPhi_3)^2
+\lambda_4 (\vPhi_1 +\vPhi_2+\vPhi_3)^4
\ee
It is clear that if we take $\lambda_i>0$ and  $\mu^2 >0$   then the vacuum corresponds to
\bea
&&|\Phi_i| = \eta^2 \label{cond1} \\
&& \vPhi_1 +\vPhi_2+\vPhi_3=0
 \label{cond2}
\eea
Note that the condition \eqref{cond2} corresponds to a
$120$ degree  configuration of the triplet of scalars, which in the antiferromagnetic   model defined in a triangular lattice corresponds to spins arranged as in the ``Mercedes-Benz'' logo.

  Concerning $V_{mix}$, the first term   is the continuum analogue of the Heisenberg interaction in antiferromagnets (the term with $\lambda_4$ coupling constant is included because  it is compatible with renormalization).

Two different ansatze were shown to lead to topologically non-trivial axially symmetric vortex-like solutions \cite{CLS}.
Written in polar coordinates they read
\begin{itemize}
\item Ansatz I:
\bea
\vPhi_1&=&f(r)(-\sin n\varphi,\cos n\varphi,0) \nonumber \\
\vPhi_2&=&f(r)(-\sin(n\varphi+\frac{2\pi}{3}),\cos(n \varphi+\frac{2\pi}{3}),0)   \nonumber\\
\vPhi_3&=& f(r)(-\sin(n \varphi+\frac{4\pi}{3}),\cos(n \varphi+\frac{4\pi}{3}),0) \nonumber\\
\vA_\varphi&=&-\frac{1}{e}(0,0,\frac{a(r)}{r})
\label{Ansatz1}
\eea
\item Ansatz II:
\ba
\hspace{1.05 cm}\vPhi_1&=&(0,0,\eta) \nonumber\\
\vPhi_2&=&\frac{1}{2}(-\sqrt{3}f(r)\sin(n\varphi),\sqrt{3}f(r)\cos(n\varphi)),-\eta)\nonumber \\
\vPhi_3&=&  \frac{1}{2}(\sqrt{3}f(r)\sin(n\varphi),-\sqrt{3}f(r)\cos(n\varphi)),-\eta) \nonumber\\
\vA_\varphi&=&-\frac{1}{e}(0,0,\frac{a(r)}{r})
\label{AA2}
\ea
\end{itemize}
with  $n \in \mathbb{Z}$. Notice that  both Ansatze  satisfy eq.~\eqref{cond2}. The conditions to ensure finite energy configurations are
\begin{align}
&\lim_{r\to 0}f(r) \sim r^{|n|} \hspace{2cm} a(0) = 0 \nonumber\\
& \lim_{r\to \infty}f(r)   = \eta  \hspace {2 cm} \lim_{r \to \infty}a(r) = -n
 \label{conditions}
\end{align}
The field equations derived from Lagrangian \eqref{L} reduce
to the radial equation
\be
 f'' +\frac{1}{r} f'-\frac{1}{r^2}(n+a)^2f = 4\lambda f(r)(f^2 - 1)
\ee
which apart from a numerical factor coincides with the radial equation for the Abelian Higgs model  equation of motion for the complex scalar if one shifts $\lambda$ according to $\lambda \to \lambda/3$ in the case of Ansatz I and $\lambda \to 8\lambda/9$ for ansatz II.

\section{The Dirac equation}
As mentioned above, inspired by the zero-mode analysis presented in \cite{CL} extending to the non-Abelian case the Jackiw-Rossi Abelian construction \cite{JR}, we shall consider the following $SU(2)$ gauge invariant Dirac Lagrangian
\be
L = \int d^3x \bar\psi
\left(i\gamma^\mu\partial_\mu \times I+ e \gamma^\mu \times A_\mu - g_a  I \times \Phi_a\right )\psi
\label{lagra}
\ee
Here $\gamma^\mu$ are the $2\times2$ gamma matrices and the background fields $A_\mu$
and $\Phi_a$ are those discussed in the previous section. Symbol $\times$ denotes
tensor product with the first factor acting in the spinorial indices and the second one
in $SU(2)$ ones.

Fermion $\psi$ is in the fundamental representation of $SU(2)$ and will be written in the form
\be
\psi = \left(
\begin{array}{cc}
\psi^U_1\\
\psi^U_2\\
\psi^D_1\\
\psi^D_2
\end{array}
\right)
\label{14}
\ee
with spinorial indices  $\rho= U,D$ and $SU(2)$ ones $f=1,2$. The fermion-scalar couplings $g_a$ have the same dimensions as the gauge coupling $e$, $[g_a] = [e] = m^{1/2}$. Note that the scalar-fermion interaction is gauge invariant.

Lagrangian \eqref{lagra} leads to the fermion field equation
\be
(i\alpha^j\partial_j \times I+ e\ \alpha^j \times A_j - g_a \ \beta \times \Phi_a)\psi = -i\partial_t \psi
\label{fir}
\ee
where $\gamma^0=\beta$, $\gamma^j=\beta \alpha^j$ and $j=1,2$ the spatial indices. We choose the Dirac matrices $\alpha^j, \beta$ in the form
\begin{align} \alpha^j &= \sigma^j \;, \;\;\;\; j=1,2 \nonumber\\ \beta &= \sigma_3
\end{align}
where $\sigma^j, \sigma^3$ are the Pauli matrices.

Following \cite{CL} we shall introduce  the transformation  $\psi \to L_3\psi$
\be
L_3 = \beta \times \sigma^3= \left(
\begin{array}{cccc}
1 & 0& 0& 0\\
0 & -1& 0& 0\\
0 & 0& -1& 0\\
0 & 0& 0& 1
\end{array}
\right)
\label{L3}
\ee
 which will play an important role in finding and classifying zero-modes which is one of the main purposes of this work.

From eq.~\eqref{fir}, the zero energy solutions satisfy
\be
 (i\sigma^j\partial_j \times 1+ e\ \sigma^j \times A_j - g_a \ \beta \times \Phi_a)\psi = 0
 \label{zm}
 \ee


To find zero mode solutions we  start by considering Ansatz I. It  will be convenient to write the gauge field   in the form
\be
A_j(\bold{r})=\frac{1}{e} \epsilon_{ji} \partial^i k(r) \sigma^3
\ee
where
\be
k = -\int_0^r \frac{a(\rho)}{\rho} d\rho
\label{19}
\ee
We now make the following
 change on the fermion field
\be
\psi(\bold{r})=T(r) X(\bold{r})
\label{decouple}
\ee
with
\begin{align}
T(r)=\exp \left(  k(r) L_3 \right)
\end{align}
so that  the gauge field in eq.~\eqref{zm} decouples and we are left with
\be
(i\sigma^j\partial_j \times I - g_a \beta \times \Phi_a) X (\bold{r}) = 0
\label{which}
\ee
The decoupling was possible because, for Ansatz I, the operator $L_3$    anticommutes with the zero mode Dirac operator in eq.~\eqref{zm}. Indeed, concerning Dirac matrices, $L_3$ anticommutes with the first two terms in eq.~\eqref{zm} and commutes with the third one while for the $SU(2)$ generators, they commute with the first two terms and anticommute with the last one. Then as a result $L_3$ anticommutes with the zero-mode Dirac operator.

 Written in components, eq.~\eqref{which} reads
\begin{align}
&(i \partial_1+\partial_2) X_1^D+  if(r) G_1 X_2^U=0 \nonumber \\
&(i \partial_1-\partial_2) X_2^U+  if(r) G_2 X_1^D=0 \label{ansL1-1} \\
&(i \partial_1+\partial_2) X_2^D-  if(r) G_2 X_1^U=0 \nonumber \\
&(i \partial_1-\partial_2) X_1^U- i f(r) G_1 X_2^D=0 \label{ansL1-3}
\end{align}
where
\begin{align}
G_1&=\left(g_1 e^{-i n \phi}+g_2 e^{-i (n \phi+2 \pi/3)}+g_3 e^{-i (n \phi+4 \pi/3)}\right)/2\nonumber \\
&=e^{-i n \phi}\left(g_1+g_2 e^{-i 2 \pi/3}+g_3 e^{-i4 \pi/3}\right)/2\equiv A e^{-i n \phi} = |A|e^{i\alpha} e^{-i n \phi}\\
G_2&=\left(g_1 e^{i n \phi}+g_2 e^{i (n \phi+2 \pi/3)}+g_3 e^{i (n \phi+4 \pi/3)}\right)/2\nonumber \\
&= e^{i n \phi}\left(g_1+g_2 e^{i 2 \pi/3}+g_3 e^{i4 \pi/3}\right)/2\equiv A^* e^{i n \phi}=
|A| e^{-i\alpha} e^{i n \phi}
\end{align}
In view of the cylindrical symmetry, it is convenient to use polar coordinates for which eqs.~\eqref{ansL1-1}-\eqref{ansL1-3} become
\begin{align}
e^{-i\phi} (i \partial_r+\frac{1}{r}\partial_\phi) X_1^D+ if(r)|A|e^{i\alpha} e^{-i n \phi} X_2^U=0 \nonumber \\
e^{i\phi} (i \partial_r-\frac{1}{r}\partial_\phi) X_2^U+  if(r) |A| e^{-i\alpha} e^{i n \phi} X_1^D=0 \label{ans1-4Fa}
\end{align}
and
\begin{align}
e^{-i\phi} (i \partial_r+\frac{1}{r}\partial_\phi) X_2^D-  if(r) |A| e^{-i\alpha} e^{i n \phi} X_1^U=0 \nonumber \\
e^{i\phi} (i \partial_r-\frac{1}{r}\partial_\phi) X_1^U - if(r) |A|e^{i\alpha} e^{-i n \phi} X_2^D=0 \label{ans1-3azi}
\end{align}

%

We now propose  for the first two equations the Ansatz
\begin{align}
X_1^D&= \chi_1^D e^{i (m-n+1) \phi}\nonumber \\
X_2^U&=\chi_2^U e^{i m \phi}
\label{first}
\end{align}
As a result, the angular dependence factorizes and the zero-mode equations for  eqs.~\eqref{ans1-4Fa} reduce to  ordinary differential  equations
\begin{align}
( \partial_r+\frac{ (m-n+1)}{r}) \chi_1^D+ f(r) \chi_2^U=0 \nonumber \\
( \partial_r-\frac{ m}{r}) \chi_2^U+ f(r) |A|^2 \chi_1^D=0 \label{ans1-2}
\end{align}
where we have redefined  $A\chi_2^U$ as $\chi_2^U$.

Similarly, for the third and forth equations we write
\begin{align}
X_1^U&=\chi_1^U e^{-i m \phi}\nonumber \\
X_2^D&= \chi_2^D e^{-i (m-n-1) \phi}
\label{third}
\end{align}
so that eqs.~\eqref{ans1-3azi} become
\begin{align}
( \partial_r-\frac{ (m-n-1)}{r}) \chi_2^D- f(r) |A|^2 \chi_1^U=0 \nonumber \\
( \partial_r+\frac{ m}{r}) \chi_1^U- f(r) \chi_2^D=0 \label{ans1-F2}
\end{align}
where we have redefined  $A\chi_2^D$ as $\chi_2^D$. Without loss of generality we choose $\chi_1^U,\chi_2^U$ and  $\chi_1^D, \chi_2^D$ real.

In view of conditions \eqref{conditions} for the vortex background, in order to have well-behaved zero-modes near the
 origin  the first set of equations, eqs.~\eqref{ans1-2}, imply
 %
\be
\begin{array}{l}
\chi_1^D \stackbin[{\rm small~r}]{\sim}{} r^{n-m-1}, r^{m+|n|+1}\\
\chi_2^U \stackbin[{\rm small~r}]{\sim}{} r^{n-m+|n|}, r^{m}
\end{array}
\label{compa2}
\ee
Compatibility of behaviors \eqref{compa2} implies
\begin{align}
n-1\ge m\ge0
\label{labeb}
\end{align}
These conditions imposes  $n$ to be a positive vortex number.

The second set of equations eqs.~\eqref{ans1-F2}, imposes
\be
\begin{array}{ll}
\chi_1^U  \stackbin[{\rm small~r}]{\sim}{} r^{m-n+|n|}, r^{-m}\\
\chi_2^D \stackbin[{\rm small~r}]{\sim}{} r^{m-n-1}, r^{-m+|n|+1}
\end{array}
\label{tible2}
\ee
Following the same procedure as above, we get in this case the following condition from eq.~\eqref{tible2}
\begin{align}
n+1\le m\le 0
\end{align}
These conditions correspond to a negative vortex number.

In summary, both for positive
and negative values of $n$ we conclude that there are $|n|$ zero modes.

Using the explicit form of $L_3$
\be
\exp(     k(r) L_3) = \left(
\begin{array}{cccc}
\exp(    k(r)) & 0 & 0 & 0\\
0 & \exp(-    k(r)) & 0 & 0\\
0 & 0 & -\exp(  k(r)) & 0\\
0 & 0 & 0 & -\exp( -  k(r))
\end{array} \right)
\ee
zero-energy eigenfunctions for Ansatz I where $n-1\ge m \ge 0$ are
\begin{align} \hspace{-0.7cm}
&\psi_{n>0}(\vec r)  =   \left(
 \begin{array}{c}
 0\\
e^{- k(r)} A^{-1}   \chi_2^U e^{i m \phi}  \\
-e^{  k(r)} \chi_1^D e^{i (m-n+1) \phi} \\
 0
 \end{array}
 \right)
 \label{ultima}
\end{align}

For the interval $n+1\le m\le 0$, the zero modes are
\begin{align} \hspace{-0.7cm}
&\psi_{n<0}(\vec r)  =    \left(
 \begin{array}{c}
e^{ k(r)}  \chi_1^U e^{-i m \phi}\\
 0\\
 0\\
 -e^{- k(r)} A^{-1}  \chi_2^D e^{-i (m-n-1) \phi}
 \end{array}
 \right)
\end{align}
Notice that the factors $\exp(\pm k(r))$ do not affect normalizability of zero modes since $k(0)=0$ and $k(r) \to \pm n \log r$ when $r \to \infty$  and the $\chi's$ are exponentially decreasing functions.

It is important to stress that $L_3$ classifies zero modes according to
\be
L_3 \psi_{n \gtrless 0}(\vec r) = \mp \psi_{n\gtrless 0}(\vec r)
\label{58}
\ee

 ~

The analysis for the case in which the background corresponds to a vortex obeying Ansatz II goes similarly. Instead of eqs.~\eqref{ansL1-3} we now have
{\begin{align}
&(i \partial_1+\partial_2) X_1^D+i f(r) H_1 e^{-i n \phi} X_2^U-H_2 X_1^U=0 \nonumber \\
&(i \partial_1-\partial_2) X_2^U
+i f(r) H_1 e^{i n \phi} X_1^D-H_2 X_2^D=0 \nonumber \\
&(i \partial_1+\partial_2) X_2^D-i f(r) H_1 e^{i n \phi} X_1^U+H_2 X_2^U=0 \nonumber \\
&(i \partial_1-\partial_2) X_1^U-i f(r) H_1 e^{- in \phi} X_2^D+H_2 X_1^D=0 \label{ans1M-4x}
\end{align}
where
\begin{align}
H_1&=\frac{\sqrt{3}}{4}(g_2-g_3)\\
H_2&=\eta \ \frac{\left(2g_1-g_2-g_3\right)}{4}
\end{align}
Notice that because of the particular form  of scalars in
Ansatz II, the presence of  the last term in the l.h.s. \!of each one of  eqs.~\eqref{ans1M-4x}
spoils the anticommutation of $L_3$  with the zero mode Dirac operator, analogously to what happens concerning chiral invariance in $3+1$ dimensions when fermions are massive. Only in the case in which these terms are absent zero-modes exist. We then impose a condition ensuring $H_2 = 0$, this implying that the following relation between coupling constants should hold
\be
2g_1-g_2-g_3 = 0
\label{condixion}
\ee
Once condition \eqref{condixion} is adopted, eqs.~\eqref{ans1M-4x} become

\begin{align}
&(i \partial_1+\partial_2) X_1^D+i f(r) H_1 e^{-i n \phi} X_2^U=0 \nonumber\\
&(i \partial_1-\partial_2) X_2^U
+i f(r) H_1 e^{i n \phi} X_1^D=0 \nonumber \\
&(i \partial_1+\partial_2) X_2^D-i f(r) H_1 e^{i n \phi} X_1^U=0 \nonumber\\
&(i \partial_1-\partial_2) X_1^U-i f(r) H_1 e^{-i n \phi} X_2^D=0 \label{ans1Mx-4}
\end{align}
or, in polar coordinates $(r,\phi)$
\begin{align}
&e^{-i\phi} (i \partial_r+\frac{1}{r}\partial_\phi) X_1^D+i f(r) H_1 e^{-i n \phi} X_2^U=0 \nonumber\\
&e^{i\phi} (i \partial_r-\frac{1}{r}\partial_\phi) X_2^U+i f(r) H_1 e^{i n \phi} X_1^D=0 \nonumber\\
&e^{-i\phi} (i \partial_r+\frac{1}{r}\partial_\phi) X_2^D-i f(r) H_1 e^{i n \phi} X_1^U=0 \nonumber\\
&e^{i\phi} (i \partial_r-\frac{1}{r}\partial_\phi) X_1^U-i f(r) H_1 e^{-i n \phi} X_2^D=0 \label{ans1-4F}
\end{align}
The adequate  phase Ansatz for $X_1^D, X_2^U$ is now
\begin{align}
X_1^D&=\chi_1^D e^{-im \phi} \nonumber\\
X_2^U&=\chi_2^U e^{i (-m+n-1) \phi}
\end{align}
leading to
\begin{align}
( \partial_r-\frac{ m}{r}) \chi_1^D+ f(r)  \chi_2^U=0 \nonumber\\
( \partial_r-\frac{ (-m+n-1)}{r}) \chi_2^U+ f(r) H_1^2 \chi_1^D=0 \label{ans1-2FS}
\end{align}
and for the other two components
\begin{align}
X_2^D& = \chi_2^D e^{im\phi}\nonumber\\
X_1^U& = \chi_1^U e^{i(m-n-1)\phi}
\end{align}
leading in this case to
\begin{align}
( \partial_r+\frac{ m}{r}) \chi_2^D- f(r)    \chi_1^U=0 \nonumber\\
( \partial_r-\frac{ (- n +m-1)}{r}) \chi_1^U- f(r) H_1^2 \chi_2^D=0 \label{ans1-4FI}
\end{align}
where we have shifted $H_1X_2^U \to X_2^U$ and $H_1 X_1^U \to X_1^U$.

  From the first set of equations we find that the appropriate behavior at the origin ensuring zero-mode regularity is
\be
\begin{array}{l}
\chi_1^D   \stackbin[{\rm small~r}]{\sim}{}  r^{m}, r^{-m+n+|n|}\\
 \chi_2^U\stackbin[{\rm small~r}]{\sim}{} r^{m+|n|+1}, r^{-m+n-1}
\end{array}
\label{compa}
\ee
and from the second,
\be
\begin{array}{ll}
\chi_1^U  \stackbin[{\rm small~r}]{\sim}{} r^{-m+|n|+1}, r^{-n+m-1} \\
\chi_2^D\stackbin[{\rm small~r}]{\sim}{} r^{-m}, r^{-n+m+|n|}
\end{array}
\label{tible}
\ee

All the solutions to these equations are regular as long as the following inequalities hold for the first set of equations
\be
n-1 \ge m\ge0
\ee
or
\be
n+1 \le m\le 0
\ee
for the second one, which are exactly the same conditions found in the case of Ansatz I. Therefore, there are also $|n|$ zero modes for Ansatz II.

The explicit form of zero-energy eigenfunctions in this case are
 \begin{align} \hspace{-0.7cm}
&\psi_{n>0}(\vec r)  =    \left(
 \begin{array}{c}
 0\\
 e^{- k(r)}H^{-1}\chi_2^U e^{i (-m+n-1) \phi}\\
 -e^{ k(r)}\chi_1^D e^{-im \phi} \\
 0
 \end{array}
 \right)
\end{align}
for positive vortex numbers. Concerning negative vortex numbers, we obtain the following zero mode
\begin{align} \hspace{-0.7cm}
&\psi_{n<0}(\vec r)  =  \left(
 \begin{array}{c}
 e^{ k(r)} H^{-1}\chi_1^U e^{i (-n+m-1) \phi}\\
 0\\
 0\\
-e^{- k(r)}  \chi_2^D e^{i m \phi}
 \end{array}
 \right)
 \label{ultimaXX}
\end{align}

Also for this Ansatz, $L_3$ classifies the zero modes as
\be
L_3 \psi_{n \gtrless 0}(\vec r) = \mp \psi_{n\gtrless 0}(\vec r)
\ee
Note that the relation between the signs of $L_3$ eigenvalues and vortex number is inverted with respect to that arising for Ansatz I, eq.~\eqref{58}.

We end this section by analyzing explicitly the only existing zero mode for the case $n = 1$. For Ansatz I it takes the form
\begin{align} \hspace{-0.7cm}
&\psi_{1}(\vec r)  =   \left(
 \begin{array}{c}
 0\\
e^{- k(r)} A^{-1} \chi_2^U \\
-e^{ k(r)}  \chi_1^D \\
 0
 \end{array}
 \right)
 \label{ultima}
\end{align}
where $ \chi_2^U$ and $\chi_1^D$ satisfy eqs.~\eqref{ans1-2}. Concerning 
 Ansatz  II we have
\begin{align} \hspace{-0.7cm}
&\psi_{1}(\vec r)  =    \left(
 \begin{array}{c}
 0\\
 e^{- k(r)}H^{-1}\chi_2^U  \\
 -e^{ k(r)}\chi_1^D   \\
 0
 \end{array}
 \right)
 \label{zeromode}
\end{align}
where $ \chi_2^U$ and $\chi_1^D$ satisfy eqs.~\eqref{compa}.

\section{Summary and discussion}

In this work we have been able to construct explicit zero modes of the Dirac equation in the gauge and scalar filds  background of the $Z_2$ vortices recently introduced in~\cite{CLS}.
We have constructed the zero modes in two different Ansatze. While for Ansatz I, zero modes exist for generic value of the scalar-fermion coupling constants, in the case of Ansatz II an explicit relation between coupling constants is required (see eq.~\eqref{condixion}).

As discussed in~\cite{CLS}, from an energetic point of view, vortices of type Ansatz II are favored against those of type Ansatz I. Also, though vortices with arbitrary $n$ are possible, their energy increases with $n$ and as a result only vortices with $|n|=1$ are topologically protected. So, the zero modes of the type given by eqs.~\eqref{ultima}-\eqref{zeromode} are
those that are relevant as well as the analogous ones with $n=-1$. One then concludes  that in this $SU(2)$ gauge invariant theory  there is only one zero mode with no angular dependence associated to a $|n| = 1$ vortex.

It can be easily seen that a simple change of the fermion basis, transforms  the Hamiltonian $H$ associated to our problem  to a Hamiltonian $\tilde H$ which is of the same type of the one considered recently by Schuster et al \cite{schuster} (see also \cite{Chamon2}-\cite{Chamon3}).
\be
{\tilde H}=\left(
\begin{array}{cccc}
0 &  -i \nabla_- - e A_- & g \Delta^*& 0 \\
 -i\nabla_+-  eA_+& 0 & 0 &  g \Delta^* \\
 g \Delta & 0 &0 &  i \nabla_- - e A_-\\
0& g \Delta & i \nabla_+ - e A_+  &0
\end{array}
\right)
\ee
where $\Delta$ is related  to the scalar fields of our Ansatze and can be written in the form $|\Delta(r)| \exp(in\phi)$.
The main difference is that our backgrounds are those arising in a non-Abelian gauge theory, and they correspond to regular solutions of finite energy.
Also, as explained before, in our non-Abelian case topology selects automatically the $|n|=1$ sector, leaving us with a single zero mode.

The vortex backgrounds considered in~\cite{CLS} were inspired by global magnetic vortices appearing in antiferromagnetic materials in the triangular lattice. In solving the zero mode problem, the gauge potential does not play a central role as it is in fact {\em decoupled} by the transformation given in eq.~\eqref{decouple}.
It would be interesting to explore in such systems if excitations coupled to the magnetization  in a similar way as in the fermion-scalar field coupling considered here do exist. Non-Abelian gauge fields also naturally arise in systems with spin-orbit interactions and cold atoms \cite{cold}. It would be interesting to analyze if nontrivial field configurations could be explicitly realized in such systems. We hope to work on these issues in a future work.

~

\noindent{\bf{Acknowledgments:}}
We would like to thank Eduardo Fradkin for helpful comments.
A.M. thanks CAPES/PNPD for the financial support. F.A.S. is associated to CICBA and financially supported by PIP-CONICET,
PICT-ANPCyT, UNLP and CICBA grants. G.S.L is finacially supported by PIP-CONICET and UBA.


\begin{thebibliography}{99}
\bibitem{JRe} R.~Jackiw and C.~Rebbi,
  Phys.\ Rev.\ D {\bf 13}, 3398 (1976).
\bibitem{thooft} G. t Hooft, Phys. Rev. Lett. {\bf 37}, 8 (1976); Phys. Rev. D {\bf 14}, 3432 (1976).
\bibitem{JR}R. Jackiw and C. Rebbi, Phys. Rev. Lett. {\bf 37}, 172 (1976).
\bibitem{Ca}C.G. Callan Jr., R.F. Dashen and D.J. Gross, Phys. Lett. B {\bf 63}, 334 (1976); Phys. Rev. D {\bf 17}, 2717 (1978).
\bibitem{Co} S. Coleman, in: The Whys of Subnuclear Physics, Ence 1977, ed. A. Zichichi (Plenum Press, New York, 1979) p. 805.
    \bibitem{JRO} R.~Jackiw and P.~Rossi,
  Nucl.\ Phys.\ B {\bf 190}, 681 (1981).
  \bibitem{We} E.~J.~Weinberg,   Phys.Rev. D {\bf 24}, 2669 (1981).
 \bibitem{Wil} F.~Wilczek, Nature {\bf 486}, 195 (2012).
 \bibitem{Witten}
  E.~Witten,
  Phys.\ Lett.\ B {\bf 153}, 243 (1985).
  \bibitem{Polchinski} See
  J.~Polchinski, hep-th/0412244 and references therein.
 \bibitem{review} For a review see for instance, M Leijense and K Flensberg, Semicond. Sci Technol. {\bf 27}, 124003 (2012), S. Das Sarma, M. Freedman, C. Nayak,npj Quantum Information 1, 15001 (2015).
 \bibitem{CL} L. F. Cugliandolo and G. Lozano, Phys.\ Rev.\ D {\bf 39}, 3093 (1989).
     \bibitem{Shifman} M.~Shifman and A.~Yung, {\it Supersymmetry Solitons}, Cambridge Univbersite Press, 2009, Cambridge, UK.
 \bibitem{GN} G.~Grignani and G.~Nardelli,   Phys.\ Rev.\ D {\bf 43}, 1919 (1991).   
\bibitem{LLM} B.-H.~Lee, C.-k.~Lee and H.~Min,  Phys.\ Rev.\ D {\bf 45}, 4588  (1992).
      \bibitem{LMS}  G.~Lozano, A.~Mohammadi and F.~A.~Schaposnik,
  JHEP {\bf 1511}, 042 (2015).
   \bibitem{Sha} M.~M.~Anber, Y.~Burnier, E.~Sabancilar and M.~Shaposhnikov,
  Phys.\ Rev.\ D {\bf 93}, 021701 (2016).
\bibitem{CLS}
  D.~Cabra, G.~S.~Lozano and F.~A.~Schaposnik,
  Phys.\ Rev.\ D {\bf 92}, 124033 (2015).
\bibitem{KM}   H.~Kawamura and  S.~Miyashita,  J.\ Phys.\ Soc.\ Jap\, {\bf 53},  4138 (1984).
\bibitem{deVS}
  H.~J.~de Vega and F.~A.~Schaposnik,
  Phys.\ Rev.\ Lett.\  {\bf 56} (1986) 2564; Phys.Rev. D {\bf 34}, 3206 (1986).
\bibitem{schuster} T.~Schuster, T.~Iadecola, C.~Chamon, R.~Jackiw and Y.S.~Pi, arXiv:1606.01905.
\bibitem{Chamon2}   C.~Chamon, C.~Y.~Hou, R.~Jackiw, C.~Mudry, S.~Y.~Pi and A.~P.~Schnyder,
  Phys.\ Rev.\ Lett.\  {\bf 100}, 110405 (2008).
  \bibitem{Chamon3}
  C.~Chamon, R.~Jackiw, Y.~Nishida, S.-Y.~Pi and L.~Santos,
  Phys.\ Rev.\ B {\bf 81}, 224515 (2010).
\bibitem{cold} K. Osterloh, M. Baig, L. Santos, P. Zoller, and M. Lewenstein
Phys.\ Rev.\ Lett. {\bf 95}, 010403 (2005).
\end{thebibliography}
\end{document}